\pdfoutput=1
\documentclass[a4paper,11pt]{article}
\usepackage{pos}
\usepackage{cleveref}

\providecommand{\repositoryInformationSetup}{} %
\repositoryInformationSetup

\usepackage{xspace}
\usepackage{bbm}
\usepackage{amsmath}
\usepackage{mathtools}
\usepackage{dsfont}
\usepackage{braket}
\usepackage{upgreek}
\usepackage[T1]{fontenc}
\usepackage[boxruled,lined,commentsnumbered]{algorithm2e}
\usepackage[utf8]{inputenc}
\usepackage[UKenglish]{babel}
\usepackage{siunitx}
\usepackage{placeins}
\usepackage{bm}
\usepackage{tikz}
\usepackage{ifthen}
\usepackage{calculator}
\usepackage{tikz-feynman}
\usepackage{graphicx}
\usepackage{slashed}
\usepackage{xfrac}
\usepackage{changes}

\graphicspath{{pics/},{figures/},{data/},{plots/},{inkscape/}}

\newcommand{\repoURL}{https://github.com/evanberkowitz/hubbard-critical-coupling}

\newcommand{\order}[1]{\ensuremath{\mathcal{O}\left(#1\right)}\xspace}

\DeclareMathOperator{\im}{i}
\DeclareMathOperator{\real}{Re}

\DeclareMathOperator{\ord}{\mathcal{O}}
\newcommand{\ordnung}[1]{\ensuremath{\ord(#1)}}

\newcommand{\erwartung}[1]{\ensuremath{\left\langle#1\right\rangle}}

\newcommand{\pdagger}{{\phantom{\dagger}}}

\newcommand{\eto}[1]{\ensuremath{\mathrm{e}^{#1}}}
\newcommand{\matr}[2]{\left(\begin{array}{#1}#2\end{array}\right)}

\let\builtinLaTeX\LaTeX
\def\LaTeX{\builtinLaTeX\xspace}

\newcommand{\grapheneColor}[2]{
	\foreach \x in {0,...,#1}
	{
		\MODULO{\x}{2}{\xmod} 
		\foreach \y in {0,...,#2}
		{
			\tikzset{yshift={\y*1.732cm+Mod(\x,2)*0.866cm}, xshift={\x*1.5cm}}
			\begin{scope}
				\draw[line width=2pt] (0,0)--(1,0);
				\ifthenelse{\xmod=0 \OR \y<#2 \AND \x>0}{\draw[line width=2pt] (0,0)--(-0.5, 0.866);}{}
				\ifthenelse{\xmod>0 \OR \y>0 \AND \x>0}{\draw[line width=2pt] (0,0)--(-0.5, -0.866);}{}
				\ifthenelse{\x=1 \AND \y=0}{\node (A1) at (0,0) {}; \node (A2) at (1.5, 0.866) {}; \node (A3) at (1.5, -0.866) {};}{}
				
				\fill[red] (0,0) circle (4pt);
				
				\fill[blue] (1,0) circle (4pt);
				\ifthenelse{\xmod=0 \OR \y<#2 \AND \x>0}{\fill[blue] (-0.5, 0.866) circle (4pt);}{}
				\ifthenelse{\y > 0 \OR \xmod > 0 \AND \x>0}{\fill[blue] (-0.5, -0.866) circle (4pt);}{}
			\end{scope}
}}}

\newcommand{\grapheneColorShift}[4]{
	\foreach \x in {0,...,#1}
	{
		\MODULO{\x}{2}{\xmod} 
		\foreach \y in {0,...,#2}
		{
			\tikzset{yshift={\y*1.732cm+Mod(\x,2)*0.866cm+#4}, xshift={\x*1.5cm+#3}}
			\begin{scope}
				\draw[line width=2pt] (0,0)--(1,0);
				\ifthenelse{\xmod=0 \OR \y<#2 \AND \x>0}{\draw[line width=2pt] (0,0)--(-0.5, 0.866);}{}
				\ifthenelse{\xmod>0 \OR \y>0 \AND \x>0}{\draw[line width=2pt] (0,0)--(-0.5, -0.866);}{}
				\ifthenelse{\x=1 \AND \y=0}{\node (A1) at (0,0) {}; \node (A2) at (1.5, 0.866) {}; \node (A3) at (1.5, -0.866) {};}{}
				
				\fill[red] (0,0) circle (4pt);
				
				\fill[blue] (1,0) circle (4pt);
				\ifthenelse{\xmod=0 \OR \y<#2 \AND \x>0}{\fill[blue] (-0.5, 0.866) circle (4pt);}{}
				\ifthenelse{\y > 0 \OR \xmod > 0 \AND \x>0}{\fill[blue] (-0.5, -0.866) circle (4pt);}{}
			\end{scope}
}}} %

\usepackage{ifthen} %
\usepackage{intcalc} %
\usepackage[skins]{tcolorbox}
\usepackage{pgfplots} %
\usepackage{grffile}  %

\usetikzlibrary{external}

\usetikzlibrary{arrows,backgrounds,fit,calc,patterns,shapes}

\usetikzlibrary{decorations.pathmorphing} %
\usetikzlibrary{intersections} %
\usetikzlibrary{positioning}	%
\usetikzlibrary{plotmarks}
\pgfplotsset{compat=newest}
\usetikzlibrary{plotmarks}
\usetikzlibrary{arrows.meta}
\usepgfplotslibrary{patchplots}

\tikzset{every node/.style={sloped,allow upside down},baseline={([yshift=+0ex]current bounding box.center)},inner sep=.3mm,x=1cm,y=1cm}

\def\pepsWidth{4mm}

\def\xDist{13mm} %
\def\yDist{10mm}
\def\xExt{\dimexpr \xDist-\pepsWidth /2 \relax} %
\def\yExt{\dimexpr \yDist-\pepsWidth /2 \relax}

\def\singularValueWidth{3mm}
\def\swapWidth{1.5mm}
\def\swapHeight{\swapWidth}
\def\gateWidth{\dimexpr \xDist+\pepsWidth \relax}
\def\gateHeight{\pepsWidth}

\tikzset{every picture/.style={node distance=\yDist and \xDist}}

\definecolor{tensorblue}{rgb}{0.8,0.8,1}
\definecolor{tensorred}{rgb}{1,0.5,0.5}
\definecolor{tensorpurple}{rgb}{1,0.5,1}

\tikzset{tenop/.style={fill=orange!30}}
\tikzset{tenk/.style={fill=green!50!black!50}}
\tikzset{tenleft/.style={fill=yellow!50}}
\tikzset{tenright/.style={fill=violet!50}}
\tikzset{tenlr/.style={fill=tensorpurple}}
\tikzset{teneff/.style={fill=tensorpurple}}
\tikzset{tentrans/.style={fill=black!20}}
\tikzset{tens/.style={fill=black!30}}

\tikzset{tenred/.style={fill=tensorred}}
\tikzset{tengreen/.style={fill=green!50!black!50}}
\tikzset{tenpurple/.style={fill=tensorpurple}}
\tikzset{tengrey/.style={fill=black!20}}
\tikzset{tenwhite/.style={fill=black!0}}
\tikzset{tenorange/.style={fill=orange!30}}
\tikzset{u/.style={fill=blue!20,draw=black}}
\tikzset{w/.style={fill=green!50!black!80,draw=black}}

\tikzset{ket/.style={circle,draw=black,thick,fill=tensorblue,minimum width=\pepsWidth}}
\tikzset{bra/.style={ket,tens,label=center:{$*$}}}
\tikzset{ketNew/.style={ket,tengreen}}
\tikzset{braNew/.style={bra,tengreen}}
\tikzset{nodeInvisible/.style={circle,draw=none,thick,minimum width=\pepsWidth}}
\tikzset{bMPS/.style={ket,tengrey}}
\tikzset{singularValue/.style={midway,circle,draw=black,thick,tenorange,minimum width=\singularValueWidth}}
\tikzset{swap/.style={diamond,draw=black,thick,fill=black,minimum width=\swapWidth,minimum height=\swapHeight}}
\tikzset{gate/.style={rectangle,draw=black,thick,fill=tensorpurple,minimum width=\gateWidth,minimum height=\gateHeight}}
\tikzset{env/.style={bMPS,rectangle,draw=black,thick,minimum height=\gateHeight}}

\tikzset{wiggly/.style={decorate, decoration={snake, segment length=\pepsWidth/2, amplitude=\pepsWidth/8}}}
\tikzset{physical/.style={densely dashed}}
\tikzset{parity/.style={dotted,very thick}}
\tikzset{boundary/.style={very thick}}

\newcommand{\rightOf}[1]{([shift=({\xDist,0})]#1.center)}

\newcommand{\belowOf}[1]{([shift=({0,-\yDist})]#1.center)}

\newcommand{\distL}[1]{([shift=({-\xExt,0})]#1.center)}
\newcommand{\distR}[1]{([shift=({\xExt,0})]#1.center)}

\newcommand{\distD}[1]{([shift=({0,-\yExt})]#1.center)}

\newcommand{\lineL}[1]{(#1) -- \distL{#1}}
\newcommand{\lineR}[1]{(#1) -- \distR{#1}}

\newcommand{\lineD}[1]{(#1) -- \distD{#1}}

\newcommand{\connectD}[2]{(#1) -- (#1|-#2.north)}

\newcommand{\gateLD}[1]{([shift=({-0.5*\xDist,-\yExt})]#1.center) -- ([shift=({-0.5*\xDist,0})]#1.center |- #1.south)}
\newcommand{\gateRD}[1]{([shift=({0.5*\xDist,-\yExt})]#1.center) -- ([shift=({0.5*\xDist,0})]#1.center |- #1.south)}

\newcommand{\critU}{\ensuremath{\num{3.84(1)}}}

\newcommand{\critNu}{\ensuremath{\num{1.18(4)}}} %

\newcommand{\critExp}{\ensuremath{\num{0.90(4)}}}

\title{Stochastic and Tensor Network simulations of the Hubbard Model}

\author*[a]{Johann Ostmeyer}

\affiliation[a]{
	Department of Mathematical Sciences,
	University of Liverpool, United Kingdom
}
 
\emailAdd{J.Ostmeyer@liverpool.ac.uk}

\abstract{
	The Hubbard model is an important tool to understand the electrical properties of various materials. More specifically, on the honeycomb lattice it is used to describe graphene predicting a quantum phase transition from a semimetal to a Mott insulating state. In this work two different numerical techniques are presented that have been employed for simulations of the Hubbard model: The Hybrid Monte Carlo algorithm on the one hand allowed us to simulate unprecedentedly large lattices, whereas Tensor Networks can be used to completely avoid the sign problem. Respective strengths and weaknesses of the methods are discussed.
}

\FullConference{%
	The 39th International Symposium on Lattice Field Theory,\\
	8th-13th August, 2022,\\
	Rheinische Friedrich-Wilhelms-Universität Bonn, Bonn, Germany
}

\makeatletter%
\begin{document}

\maketitle

\section{Introduction 
\label{sec:intro}}

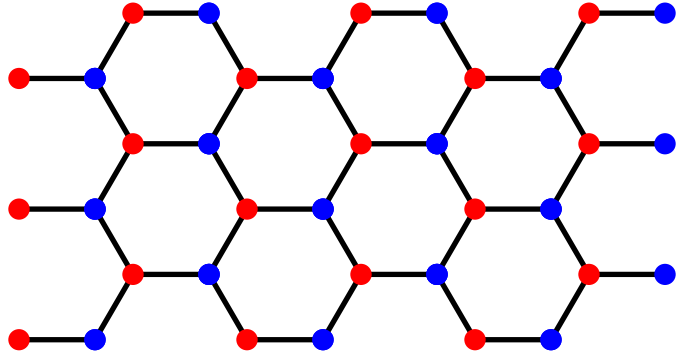
\begin{figure}[ht]
	\centering
	\begin{tikzpicture}
		\grapheneColor{5}{2}
	\end{tikzpicture}
	\caption[Honeycomb lattice of graphene.]{Honeycomb lattice of graphene. The red and the blue points form the two triangular sublattices respectively.}
	\label{fig:honeycomb_lattice}
\end{figure}

Graphene is the only known material consisting of a single atomic layer~\cite{GeimNovoselovReview}. Carbon atoms form a honeycomb lattice consisting out of two triangular Bravais sublattices with each site's nearest neighbours belonging to the opposite sublattice as shown in \cref{fig:honeycomb_lattice}. This means that the lattice can be coloured using two alternating colours. Graphene and derived carbon nanostructures like nanotubes and fullerenes have unique electromagnetic properties~\cite{CastroNeto2009}.

In order investigate these properties theoretically, we employ the Hubbard model which describes electronic interactions in a simple way. It is assumed that the carbon atoms composing graphene have fixed lattice positions and moreover not more than two electrons per site are allowed to move and thus contribute to the electromagnetic properties of the material. These electrons are confined to the lattice points at any given time, but they can instantly hop from one lattice point to a nearest neighbour. Hence, exactly zero, one or two electrons (of opposite spin) can be at the same lattice point simultaneously. In addition, an on-site interaction $U$ models the repulsive force of the identically charged particles and a chemical potential $\mu$ governs the total electron number.

We use a particle-hole basis~\cite{Luu:2015gpl}, that is we count the present spin-up particles and the absent spin-down particles, therefore our Hamiltonian reads
\begin{equation}
	H=-\kappa \sum_{\erwartung{x,y}}\left(p^\dagger_{x}p^\pdagger_{y}+h^\dagger_{x}h^\pdagger_{y}\right)+\frac{U}{2}\sum_{x}\rho_x\rho_x+\mu\sum_x\rho_x\,,\qquad \rho_x = p^\dagger_x p_x -h^\dagger_x h_x\,,\label{eqn:article_hole_hamiltonian}
\end{equation}
where $\erwartung{x,y}$ denotes nearest neighbour tuples, $p$ and $h$ are fermionic particle and hole annihilation operators, $\kappa$ is the hopping amplitude and $\rho_x$ is the charge operator.

\begin{figure}[ht]
\renewcommand{\grapheneColorShift}[4]{
	\foreach \x in {0,...,#1}
	{
		\MODULO{\x}{2}{\xmod} 
		\foreach \y in {0,...,#2}
		{
			\tikzset{yshift={\y*1.732cm+Mod(\x,2)*0.866cm+#4}, xshift={\x*1.5cm+#3}}
			\begin{scope}
				\draw[line width=2pt] (0,0)--(1,0);
				\ifthenelse{\xmod=0 \OR \y<#2 \AND \x>0}{\draw[line width=2pt] (0,0)--(-0.5, 0.866);}{}
				\ifthenelse{\xmod>0 \OR \y>0 \AND \x>0}{\draw[line width=2pt] (0,0)--(-0.5, -0.866);}{}
				\ifthenelse{\x=1 \AND \y=0}{\node (A1) at (0,0) {}; \node (A2) at (1.5, 0.866) {}; \node (A3) at (1.5, -0.866) {};}{}
				
				\fill[red] (0,0) circle (4pt);
				
				\fill[blue] (1,0) circle (4pt);
				\ifthenelse{\xmod=0 \OR \y<#2 \AND \x>0}{\fill[blue] (-0.5, 0.866) circle (4pt);}{}
				\ifthenelse{\y > 0 \OR \xmod > 0 \AND \x>0}{\fill[blue] (-0.5, -0.866) circle (4pt);}{}
			\end{scope}
}}}

\hfill
\begin{tikzpicture}
	\node[anchor=south west,inner sep=0] (band) at (0,-1) {\includegraphics[height=0.25\textheight]{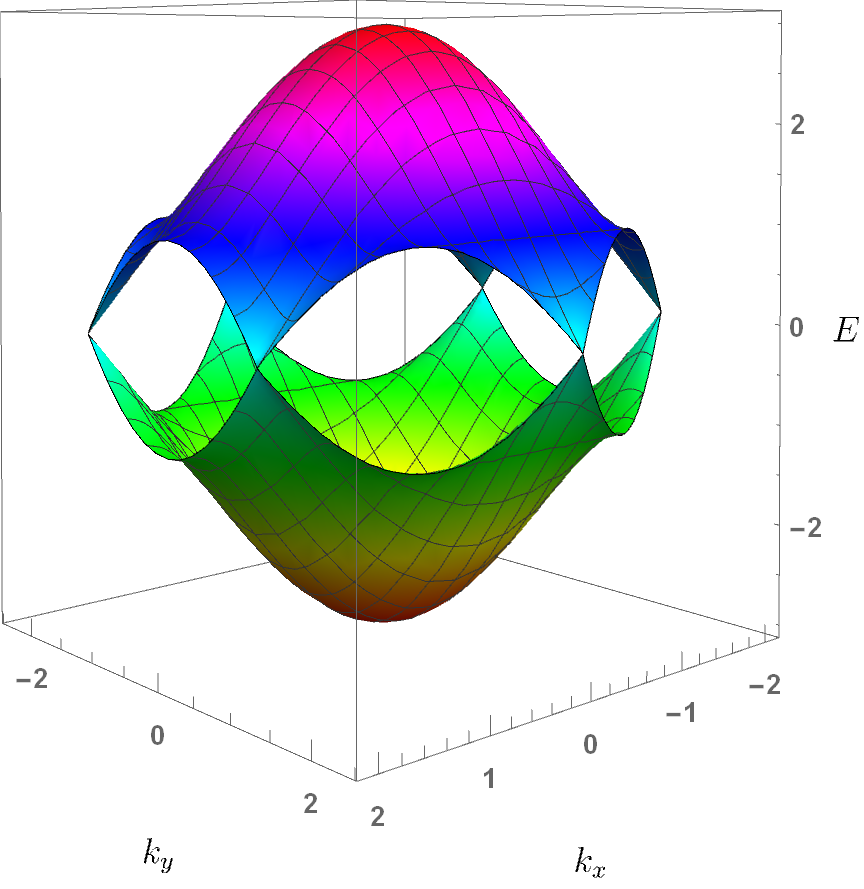}};
	\node[anchor=south west,inner sep=0] (cone) at (6.4,2.4) {\includegraphics[height=0.11\textheight]{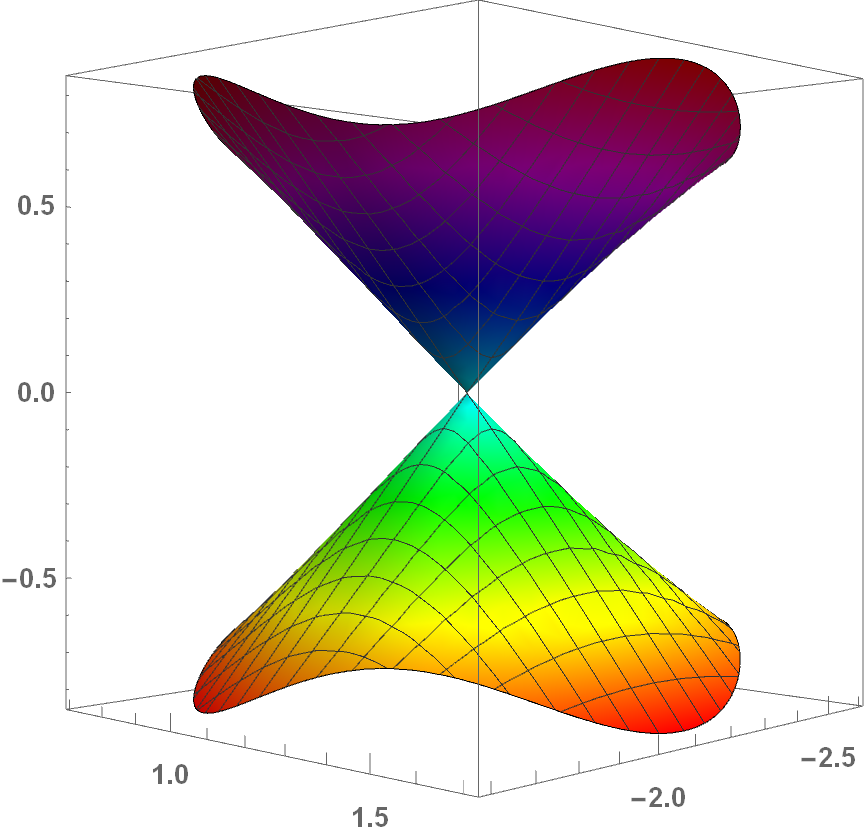}};
	\node[anchor=south west,inner sep=0] (open) at (6.4,-1.) {\includegraphics[height=0.11\textheight]{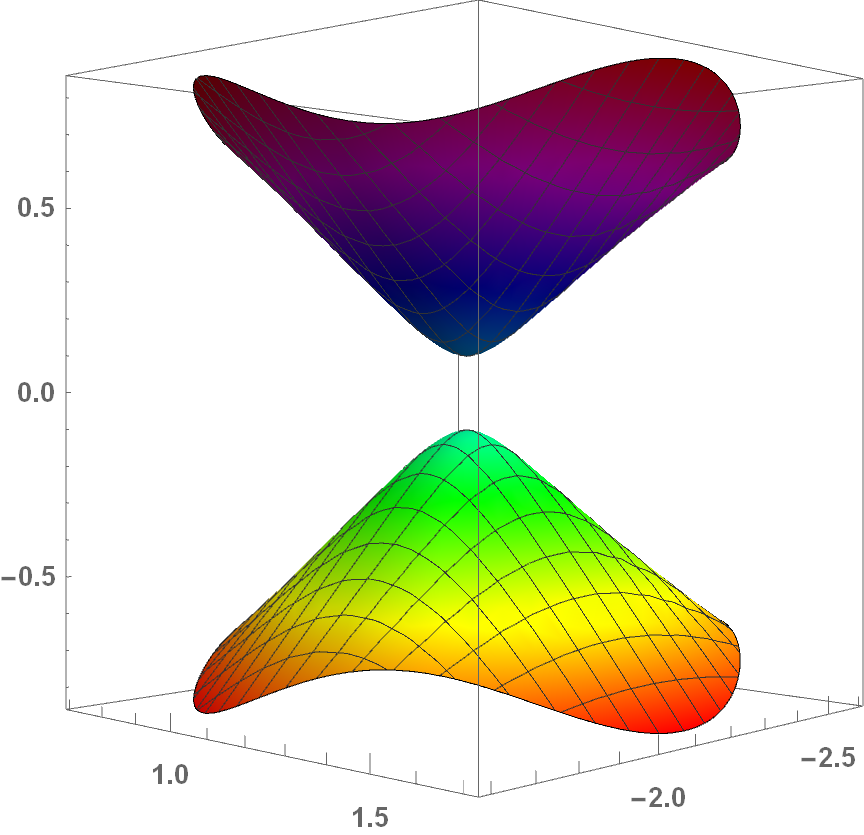}};
	\node at (8,0.3) {\footnotesize$\}$\Large};
	\node at (8.2,0.3) {$\Delta$};
	\begin{scope}[x={(cone.south east)},y={(cone.north west)}]
		\draw[->,ultra thick] (cone) -- (open) node[midway,right,rotate=90] {$U>U_c$};
	\end{scope}
	\color{black}
	\begin{scope}
		\draw[thick] (4.2,2.7) circle (0.35);
		\draw[thick] (4.2,3.05) -- (6.6,4.6);
		\draw[thick] (4.2,2.35) -- (7.58,2.57);
	\end{scope}
\end{tikzpicture}
\hfill
\hfill
\raisebox{0.1\height}{
	\begin{tikzpicture}
		\grapheneColor{0}{0}
		\draw[thick,->] (0,.3)--(0,1);
		\draw[thick,->] (1,1)--(1,.3);
	\node at (1.5,0) {$+$};		
		\grapheneColorShift{0}{0}{2cm}{0}
		\draw[thick,->] (2,1)--(2,.3);
		\draw[thick,->] (3,.3)--(3,1);
		
		\node at (-.5,-1.5) {$+$};		
		\grapheneColorShift{0}{0}{0}{-1.5cm}
		\draw[thick,->] (-0.1,-.5)--(-0.1,-1.2);
		\draw[thick,->] (.1,-1.2)--(.1,-.5);
		
		\node at (1.5,-1.5) {$+$};		
		\grapheneColorShift{0}{0}{2cm}{-1.5cm}
		\draw[thick,->] (2.9,-.5)--(2.9,-1.2);
		\draw[thick,->] (3.1,-1.2)--(3.1,-.5);
		
		\draw[->,ultra thick] (1.,-2) -- (.5,-2.7) node[midway,right,rotate=125.2] {$\;U>U_c$};
		\draw[->,ultra thick] (2.,-2) -- (2.5,-2.7);
		
		\grapheneColorShift{0}{0}{0}{-4cm}
		\draw[thick,->] (0,-3.7)--(0,-3);
		\draw[thick,->] (1,-3)--(1,-3.7);
		\node at (1.5,-4) {or};		
		\grapheneColorShift{0}{0}{2cm}{-4cm}
		\draw[thick,->] (2,-3)--(2,-3.7);
		\draw[thick,->] (3,-3.7)--(3,-3);
	\end{tikzpicture}
}
\hfill 	\caption{Left: The two energy bands (in multiples of the hopping $\kappa$) of the non-interacting Hubbard model as a function of the momentum $k$ normalised by the lattice spacing $a$. Center: Inset showing the Dirac cones. A band gap $\Delta$ separating the bands opens in the phase transition, once a critical coupling $U_c$ is surpassed. The bottom figure is only a qualitative visualisation, not the exact result. Right: The sublattice symmetry is broken at the same critical coupling and the disordered state (a superposition of all possibilities) transitions to an antiferromagnetic order.}
	\label{fig:opening_gap}
\end{figure}
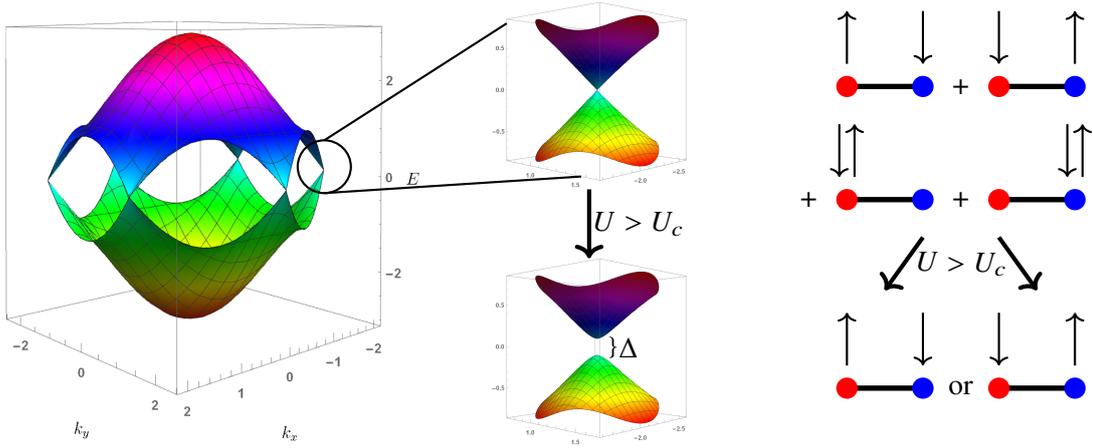

There are special cases in which the Hubbard model on the honeycomb lattice can be solved exactly. For instance the tight binding limit with $U=0$ has an analytic solution that features two energy bands touching at the so called Dirac points with a linear (relativistic) dispersion relation as depicted on the left in \cref{fig:opening_gap}. Furthermore the density of states goes to zero at exactly this point. These two properties define a semimetal and they are in surprisingly good agreement with experimental measurements of graphene which is found to be a good electric conductor.
In contrast to the hopping strength $\kappa\approx \SI{2.7}{\electronvolt}$ well determined experimentally for graphene, the coupling $U$ is not known from experiment.
Moreover the general Hubbard model with $U\neq 0$ has neither analytic nor perturbative solutions and exact numerical solutions become unfeasible for physically interesting numbers of lattice sites because the dimension of the Fock space grows exponentially with the lattice size. This necessitates approximate solutions like the stochastic and tensor network algorithms we introduce below.

By now it is well known that the Hubbard model on the honeycomb lattice undergoes a zero-temperature quantum phase transition at some critical coupling $U_c$~\cite{Assaad:2013xua,Otsuka:2015iba}. For $U<U_c$ the system is in a conducting semi-metallic state, while above this critical coupling a band gap opens (visualised in the central column of \cref{fig:opening_gap}), so it becomes a Mott insulator.
Experimentally, the value of $U$ in graphene can be confined to the region $U<U_c$ without Mott gap~\cite{Kotov2012}, the value of $U_c$ however cannot be measured. $U_c$ therefore has to be determined by theoretical or numerical investigations of the Hubbard model as we do in this work.

It has also been established for some time that an antiferromagnetic (AFM) order is formed in the insulating state (see fig.~\ref{fig:opening_gap}, right) and we could recently show~\cite{more_observables} that both, insulating and AFM, transitions happen simultaneously.

The rest of this proceeding is structured as follows. In \Cref{sec:hmc} we explain how the Hybrid Monte-Carlo (HMC) algorithm allowed us to simulate unprecedentedly large honeycomb lattices at half filling ($\mu=0$) and to analyse the quantum phase transition to a high precision. The most important physical results are summarised as well. Next, in \Cref{sec:peps}, we introduce the sign problem that occurs away from half filling ($\mu\neq0$) and we show how it can be overcome with the use of Tensor Networks (TN), but we also address the limitations this approach has so far. Finally, a comparison of the two approaches is provided in \Cref{sec:conclusion}. Advantages and disadvantages of HMC and TN algorithms respectively are discussed.

\section{Hybrid Monte Carlo}\label{sec:hmc}

Numerous approaches have been utilised to solve the Hubbard model.
The majority of algorithms dealing with the Hubbard model at half filling (or at small chemical potential) belong to the class of quantum Monte Carlo (QMC) simulations. Stochastic simulations arise naturally from the probabilistic nature of quantum mechanics and they have proven to be very successful.
In this work we use the HMC algorithm, a Markov-chain Monte Carlo (MCMC) method with global updates on continuous fields. A simple pedagogical introduction to the HMC algorithm can be found in~\cite{my_ising}.
Brower, Rebbi and Schaich (BRS) originally proposed to use the HMC algorithm for simulations of graphene~\cite{Brower:2012zd}. Their formalism stands in stark contrast to the widespread local Blankenbecler-Sugar-Scalapino (BSS)~\cite{Blankenbecler:1981jt} algorithm.
The main advantage of the HMC over local update schemes like the BSS algorithm is its superior scaling with volume $\order{V^{5/4}}$ whereas most alternative schemes scale as volume cubed $\order{V^3}$.
In practice BSS usually outperforms BRS on small systems where it is less noisy, but the HMC (i.e.\ BRS) gains the upper hand on large lattices which are essential for approaching the thermodynamic limit.
In addition, the HMC has been heavily optimised, in particular in lattice quantum chromodynamics (QCD), and we utilised many of these improvements for our condensed matter simulations~\cite{acceleratingHMC}.

By the time this work started, HMC simulations of the Hubbard model had been well established~\cite{Smith:2014tha,Luu:2015gpl,Wynen:2018ryx,Buividovich:2018yar}.
The algorithmic details at half filling including our optimisation methods are to be found in~\cite{acceleratingHMC}.
In short, we formulate the problem on a lattice in 2+1 Euclidean dimensions at finite inverse temperature $\beta$ and perform a Hubbard-Stratonovich transformation in order to obtain the effective Hamiltonian
\begin{equation}
	\mathcal{H}=\frac{\delta}{2 U} \phi^2 +\chi^\dagger\left(MM^\dagger\right)^{-1}\!\chi +\frac{1}{2}\pi^2\label{hamiltonian}
\end{equation}
Here $\pi$ is the real momentum field, $\phi$ the real Hubbard field, $\chi$ a complex pseudofermionic vector field, $\delta=\beta/N_t$ is the time step size, and $M$ is the fermion operator with
\begin{equation}
	\begin{split}
		M^{AA}_{(x,t)(y,t')}&=\delta_{xy}\left(\delta_{t+1,t'}-\delta_{tt'}\eto{\im\delta\phi_{x,t}}\right)\\
		M^{BB}_{(x,t)(y,t')}&=\delta_{xy}\left(\delta_{tt'}-\delta_{t-1,t'}\eto{-\im\delta\phi_{x,t}}\right)\\
		M^{AB}_{(x,t)(y,t')}=M^{BA}_{(x,t)(y,t')}&=-\delta\kappa\delta_{\erwartung{x,y}}\delta_{t,t'}\,.
	\end{split}\label{ferm_op}
\end{equation}
The HMC algorithm now generates $\pi$ and an auxiliary complex field $\rho$ according to a gaussian distribution $\eto{-\pi^2/2}$ respectively $\eto{-\rho^\dagger\rho}$. Then the pseudofermionic field is obtained as $\chi=M\rho$. With this starting parameters and an initial field $\phi$ a molecular dynamics trajectory is calculated and the result is accepted with the probability $\text{min}\!\left(1,\eto{-\Delta \mathcal{H}}\right)$. $\Delta \mathcal{H}$ is the difference in energy resulting from the molecular dynamics. This procedure guarantees sampling according to the probability density $p[\phi]\propto\det\left(MM^\dagger\right)\eto{-\frac{\delta}{2 U} \phi^2}$.

Our optimised methods allowed for the largest lattices simulated to date (20,808 lattice sites) enabling us to perform the first thorough analysis and elimination of all finite size and discretisation effects.
The data analysis, in particular a high number of plateau fits~\cite{prony_gevm}, has been performed using the \texttt{hadron} package~\cite{hadron} in \texttt{R}~\cite{R_language}.
We calculated the single particle gap $\Delta$ and the staggered magnetisation $m_s$ as order parameters of the conductor-insulator~\cite{semimetalmott} and the AFM~\cite{more_observables} transitions respectively. In both cases a data collapse onto a universal function allowed to extract the critical coupling $U_c/\kappa=\critU$ as well as the critical exponents $\nu=\critNu$, $\upbeta=\critExp$, and $\eta_\phi=\num{0.52(1)}$.

\begin{figure}[t]
	\raisebox{0.06\height}{
		\resizebox{0.5\textwidth}{!}{{\Large
\begingroup
  \inputencoding{latin1}%
  \makeatletter
  \providecommand\color[2][]{%
    \GenericError{(gnuplot) \space\space\space\@spaces}{%
      Package color not loaded in conjunction with
      terminal option `colourtext'%
    }{See the gnuplot documentation for explanation.%
    }{Either use 'blacktext' in gnuplot or load the package
      color.sty in LaTeX.}%
    \renewcommand\color[2][]{}%
  }%
  \providecommand\includegraphics[2][]{%
    \GenericError{(gnuplot) \space\space\space\@spaces}{%
      Package graphicx or graphics not loaded%
    }{See the gnuplot documentation for explanation.%
    }{The gnuplot epslatex terminal needs graphicx.sty or graphics.sty.}%
    \renewcommand\includegraphics[2][]{}%
  }%
  \providecommand\rotatebox[2]{#2}%
  \@ifundefined{ifGPcolor}{%
    \newif\ifGPcolor
    \GPcolortrue
  }{}%
  \@ifundefined{ifGPblacktext}{%
    \newif\ifGPblacktext
    \GPblacktexttrue
  }{}%
  \let\gplgaddtomacro\g@addto@macro
  \gdef\gplbacktext{}%
  \gdef\gplfronttext{}%
  \makeatother
  \ifGPblacktext
    \def\colorrgb#1{}%
    \def\colorgray#1{}%
  \else
    \ifGPcolor
      \def\colorrgb#1{\color[rgb]{#1}}%
      \def\colorgray#1{\color[gray]{#1}}%
      \expandafter\def\csname LTw\endcsname{\color{white}}%
      \expandafter\def\csname LTb\endcsname{\color{black}}%
      \expandafter\def\csname LTa\endcsname{\color{black}}%
      \expandafter\def\csname LT0\endcsname{\color[rgb]{1,0,0}}%
      \expandafter\def\csname LT1\endcsname{\color[rgb]{0,1,0}}%
      \expandafter\def\csname LT2\endcsname{\color[rgb]{0,0,1}}%
      \expandafter\def\csname LT3\endcsname{\color[rgb]{1,0,1}}%
      \expandafter\def\csname LT4\endcsname{\color[rgb]{0,1,1}}%
      \expandafter\def\csname LT5\endcsname{\color[rgb]{1,1,0}}%
      \expandafter\def\csname LT6\endcsname{\color[rgb]{0,0,0}}%
      \expandafter\def\csname LT7\endcsname{\color[rgb]{1,0.3,0}}%
      \expandafter\def\csname LT8\endcsname{\color[rgb]{0.5,0.5,0.5}}%
    \else
      \def\colorrgb#1{\color{black}}%
      \def\colorgray#1{\color[gray]{#1}}%
      \expandafter\def\csname LTw\endcsname{\color{white}}%
      \expandafter\def\csname LTb\endcsname{\color{black}}%
      \expandafter\def\csname LTa\endcsname{\color{black}}%
      \expandafter\def\csname LT0\endcsname{\color{black}}%
      \expandafter\def\csname LT1\endcsname{\color{black}}%
      \expandafter\def\csname LT2\endcsname{\color{black}}%
      \expandafter\def\csname LT3\endcsname{\color{black}}%
      \expandafter\def\csname LT4\endcsname{\color{black}}%
      \expandafter\def\csname LT5\endcsname{\color{black}}%
      \expandafter\def\csname LT6\endcsname{\color{black}}%
      \expandafter\def\csname LT7\endcsname{\color{black}}%
      \expandafter\def\csname LT8\endcsname{\color{black}}%
    \fi
  \fi
    \setlength{\unitlength}{0.0500bp}%
    \ifx\gptboxheight\undefined%
      \newlength{\gptboxheight}%
      \newlength{\gptboxwidth}%
      \newsavebox{\gptboxtext}%
    \fi%
    \setlength{\fboxrule}{0.5pt}%
    \setlength{\fboxsep}{1pt}%
\begin{picture}(7200.00,5040.00)%
    \gplgaddtomacro\gplbacktext{%
      \csname LTb\endcsname%
      \put(814,921){\makebox(0,0)[r]{\strut{}$0$}}%
      \csname LTb\endcsname%
      \put(814,1354){\makebox(0,0)[r]{\strut{}$0.1$}}%
      \csname LTb\endcsname%
      \put(814,1787){\makebox(0,0)[r]{\strut{}$0.2$}}%
      \csname LTb\endcsname%
      \put(814,2220){\makebox(0,0)[r]{\strut{}$0.3$}}%
      \csname LTb\endcsname%
      \put(814,2653){\makebox(0,0)[r]{\strut{}$0.4$}}%
      \csname LTb\endcsname%
      \put(814,3086){\makebox(0,0)[r]{\strut{}$0.5$}}%
      \csname LTb\endcsname%
      \put(814,3520){\makebox(0,0)[r]{\strut{}$0.6$}}%
      \csname LTb\endcsname%
      \put(814,3953){\makebox(0,0)[r]{\strut{}$0.7$}}%
      \csname LTb\endcsname%
      \put(814,4386){\makebox(0,0)[r]{\strut{}$0.8$}}%
      \csname LTb\endcsname%
      \put(814,4819){\makebox(0,0)[r]{\strut{}$0.9$}}%
      \csname LTb\endcsname%
      \put(1246,484){\makebox(0,0){\strut{}$1$}}%
      \csname LTb\endcsname%
      \put(1997,484){\makebox(0,0){\strut{}$1.5$}}%
      \csname LTb\endcsname%
      \put(2748,484){\makebox(0,0){\strut{}$2$}}%
      \csname LTb\endcsname%
      \put(3499,484){\makebox(0,0){\strut{}$2.5$}}%
      \csname LTb\endcsname%
      \put(4250,484){\makebox(0,0){\strut{}$3$}}%
      \csname LTb\endcsname%
      \put(5001,484){\makebox(0,0){\strut{}$3.5$}}%
      \csname LTb\endcsname%
      \put(5500,484){\makebox(0,0){\strut{}$U_c$}}%
      \csname LTb\endcsname%
      \put(5752,484){\makebox(0,0){\strut{}$4$}}%
      \csname LTb\endcsname%
      \put(6503,484){\makebox(0,0){\strut{}$4.5$}}%
    }%
    \gplgaddtomacro\gplfronttext{%
      \csname LTb\endcsname%
      \put(198,2761){\rotatebox{-270}{\makebox(0,0){\strut{}$\Delta$}}}%
      \put(3874,154){\makebox(0,0){\strut{}$U$}}%
      \csname LTb\endcsname%
      \put(1870,4646){\makebox(0,0)[r]{\strut{}$\beta = 3$}}%
      \csname LTb\endcsname%
      \put(1870,4426){\makebox(0,0)[r]{\strut{}$\beta = 4$}}%
      \csname LTb\endcsname%
      \put(1870,4206){\makebox(0,0)[r]{\strut{}$\beta = 6$}}%
      \csname LTb\endcsname%
      \put(1870,3986){\makebox(0,0)[r]{\strut{}$\beta = 8$}}%
      \csname LTb\endcsname%
      \put(1870,3766){\makebox(0,0)[r]{\strut{}$\beta = 10$}}%
      \csname LTb\endcsname%
      \put(1870,3546){\makebox(0,0)[r]{\strut{}$\beta = 12$}}%
      \csname LTb\endcsname%
      \put(1870,3326){\makebox(0,0)[r]{\strut{}$\beta = \infty$}}%
    }%
    \gplbacktext
    \put(0,0){\includegraphics{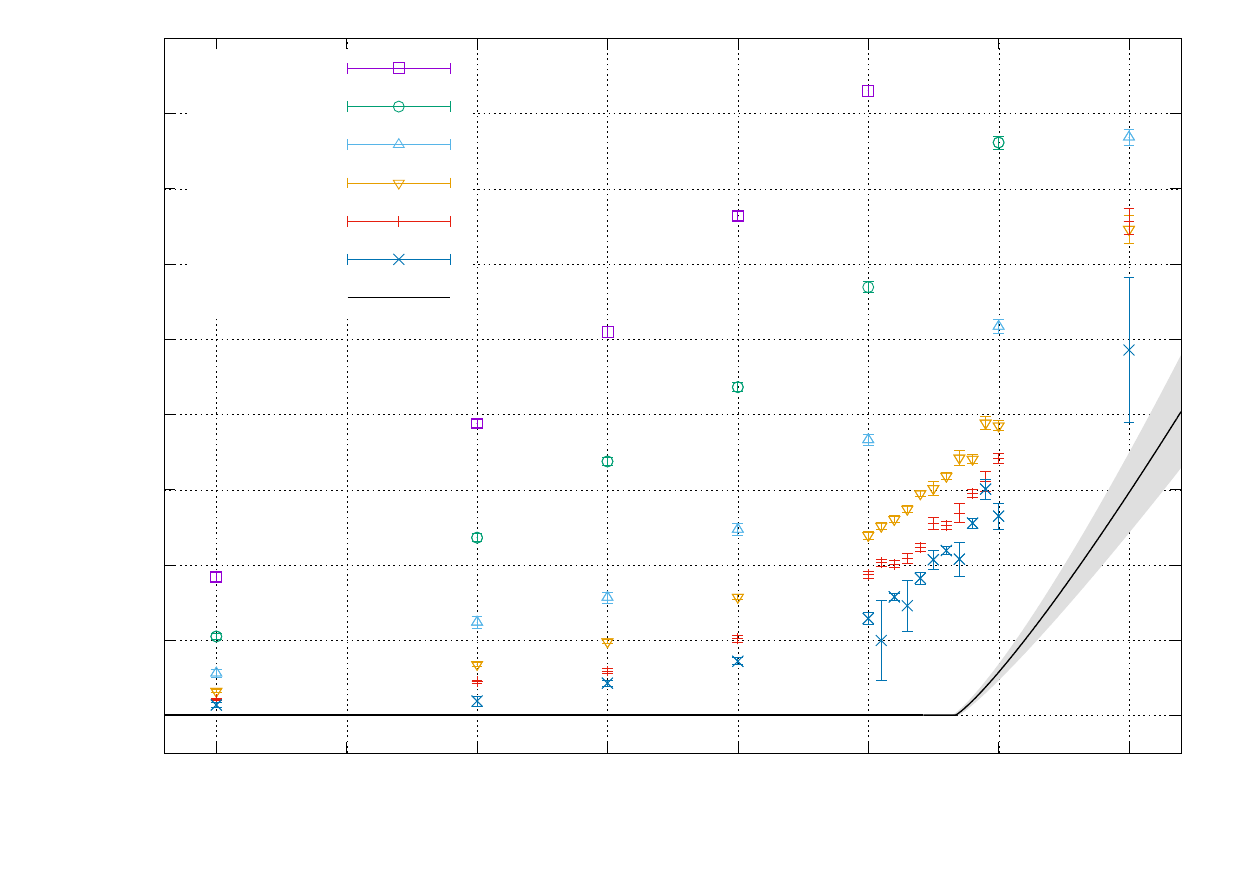}}%
    \gplfronttext
  \end{picture}%
\endgroup
}}}
	\resizebox{0.505\textwidth}{!}{{%
\begingroup
  \inputencoding{latin1}%
  \makeatletter
  \providecommand\color[2][]{%
    \GenericError{(gnuplot) \space\space\space\@spaces}{%
      Package color not loaded in conjunction with
      terminal option `colourtext'%
    }{See the gnuplot documentation for explanation.%
    }{Either use 'blacktext' in gnuplot or load the package
      color.sty in LaTeX.}%
    \renewcommand\color[2][]{}%
  }%
  \providecommand\includegraphics[2][]{%
    \GenericError{(gnuplot) \space\space\space\@spaces}{%
      Package graphicx or graphics not loaded%
    }{See the gnuplot documentation for explanation.%
    }{The gnuplot epslatex terminal needs graphicx.sty or graphics.sty.}%
    \renewcommand\includegraphics[2][]{}%
  }%
  \providecommand\rotatebox[2]{#2}%
  \@ifundefined{ifGPcolor}{%
    \newif\ifGPcolor
    \GPcolortrue
  }{}%
  \@ifundefined{ifGPblacktext}{%
    \newif\ifGPblacktext
    \GPblacktexttrue
  }{}%
  \let\gplgaddtomacro\g@addto@macro
  \gdef\gplbacktext{}%
  \gdef\gplfronttext{}%
  \makeatother
  \ifGPblacktext
    \def\colorrgb#1{}%
    \def\colorgray#1{}%
  \else
    \ifGPcolor
      \def\colorrgb#1{\color[rgb]{#1}}%
      \def\colorgray#1{\color[gray]{#1}}%
      \expandafter\def\csname LTw\endcsname{\color{white}}%
      \expandafter\def\csname LTb\endcsname{\color{black}}%
      \expandafter\def\csname LTa\endcsname{\color{black}}%
      \expandafter\def\csname LT0\endcsname{\color[rgb]{1,0,0}}%
      \expandafter\def\csname LT1\endcsname{\color[rgb]{0,1,0}}%
      \expandafter\def\csname LT2\endcsname{\color[rgb]{0,0,1}}%
      \expandafter\def\csname LT3\endcsname{\color[rgb]{1,0,1}}%
      \expandafter\def\csname LT4\endcsname{\color[rgb]{0,1,1}}%
      \expandafter\def\csname LT5\endcsname{\color[rgb]{1,1,0}}%
      \expandafter\def\csname LT6\endcsname{\color[rgb]{0,0,0}}%
      \expandafter\def\csname LT7\endcsname{\color[rgb]{1,0.3,0}}%
      \expandafter\def\csname LT8\endcsname{\color[rgb]{0.5,0.5,0.5}}%
    \else
      \def\colorrgb#1{\color{black}}%
      \def\colorgray#1{\color[gray]{#1}}%
      \expandafter\def\csname LTw\endcsname{\color{white}}%
      \expandafter\def\csname LTb\endcsname{\color{black}}%
      \expandafter\def\csname LTa\endcsname{\color{black}}%
      \expandafter\def\csname LT0\endcsname{\color{black}}%
      \expandafter\def\csname LT1\endcsname{\color{black}}%
      \expandafter\def\csname LT2\endcsname{\color{black}}%
      \expandafter\def\csname LT3\endcsname{\color{black}}%
      \expandafter\def\csname LT4\endcsname{\color{black}}%
      \expandafter\def\csname LT5\endcsname{\color{black}}%
      \expandafter\def\csname LT6\endcsname{\color{black}}%
      \expandafter\def\csname LT7\endcsname{\color{black}}%
      \expandafter\def\csname LT8\endcsname{\color{black}}%
    \fi
  \fi
    \setlength{\unitlength}{0.0500bp}%
    \ifx\gptboxheight\undefined%
      \newlength{\gptboxheight}%
      \newlength{\gptboxwidth}%
      \newsavebox{\gptboxtext}%
    \fi%
    \setlength{\fboxrule}{0.5pt}%
    \setlength{\fboxsep}{1pt}%
\begin{picture}(5040.00,3772.00)%
    \gplgaddtomacro\gplbacktext{%
      \csname LTb\endcsname%
      \put(814,704){\makebox(0,0)[r]{\strut{}$0$}}%
      \csname LTb\endcsname%
      \put(814,1111){\makebox(0,0)[r]{\strut{}$0.1$}}%
      \csname LTb\endcsname%
      \put(814,1517){\makebox(0,0)[r]{\strut{}$0.2$}}%
      \csname LTb\endcsname%
      \put(814,1924){\makebox(0,0)[r]{\strut{}$0.3$}}%
      \csname LTb\endcsname%
      \put(814,2331){\makebox(0,0)[r]{\strut{}$0.4$}}%
      \csname LTb\endcsname%
      \put(814,2738){\makebox(0,0)[r]{\strut{}$0.5$}}%
      \csname LTb\endcsname%
      \put(814,3144){\makebox(0,0)[r]{\strut{}$0.6$}}%
      \csname LTb\endcsname%
      \put(814,3551){\makebox(0,0)[r]{\strut{}$0.7$}}%
      \csname LTb\endcsname%
      \put(3274,484){\makebox(0,0){\strut{}$U_c$}}%
      \csname LTb\endcsname%
      \put(3411,484){\makebox(0,0){\strut{}}}%
      \csname LTb\endcsname%
      \put(946,484){\makebox(0,0){\strut{}$1$}}%
      \csname LTb\endcsname%
      \put(1357,484){\makebox(0,0){\strut{}$1.5$}}%
      \csname LTb\endcsname%
      \put(1768,484){\makebox(0,0){\strut{}$2$}}%
      \csname LTb\endcsname%
      \put(2178,484){\makebox(0,0){\strut{}$2.5$}}%
      \csname LTb\endcsname%
      \put(2589,484){\makebox(0,0){\strut{}$3$}}%
      \csname LTb\endcsname%
      \put(3000,484){\makebox(0,0){\strut{}$3.5$}}%
      \csname LTb\endcsname%
      \put(3821,484){\makebox(0,0){\strut{}$4.5$}}%
      \csname LTb\endcsname%
      \put(4232,484){\makebox(0,0){\strut{}$5$}}%
      \csname LTb\endcsname%
      \put(4643,484){\makebox(0,0){\strut{}$5.5$}}%
    }%
    \gplgaddtomacro\gplfronttext{%
      \csname LTb\endcsname%
      \put(209,2127){\rotatebox{-270}{\makebox(0,0){\strut{}$m_s$}}}%
      \put(2794,154){\makebox(0,0){\strut{}$U$}}%
    }%
    \gplbacktext
    \put(0,0){\includegraphics{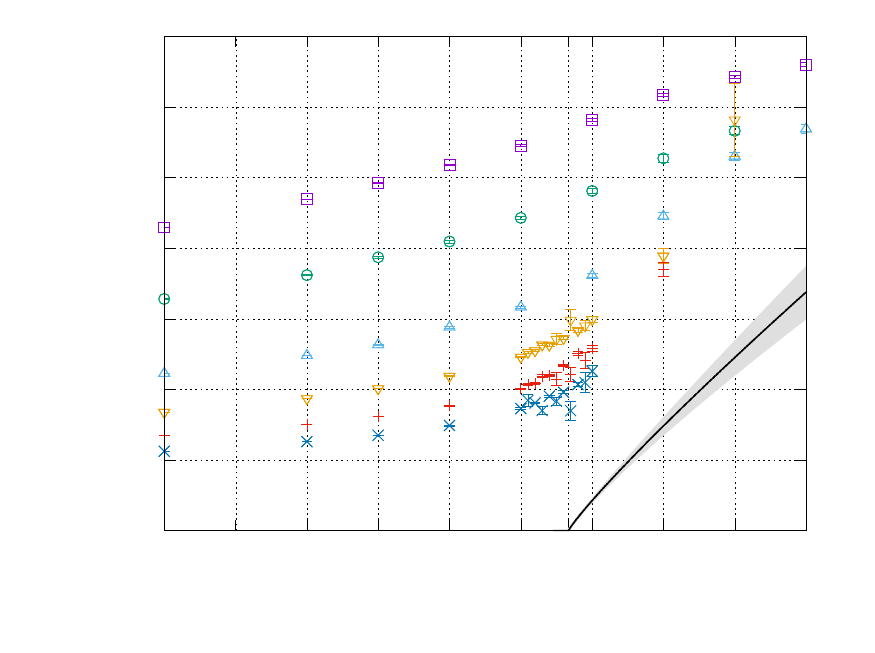}}%
    \gplfronttext
  \end{picture}%
\endgroup
}}
	\caption{All quantities in units of $\kappa$ and after the thermodynamic and continuum limit extrapolations. $\beta$ is the inverse temperature.
		The single-particle gap $\Delta(U,\beta)$ (left) and the AFMI order parameter (staggered magnetization) $m_s$ (right).  
		We also show $\Delta(U,\beta = \infty)$ and $m_s(U,\beta = \infty)$ as solid black lines with error band. The legend from the left plot applies to both.}
	\label{fig:both_order_parameters}
\end{figure}

\Cref{fig:both_order_parameters} shows the order parameters.
In the zero-temperature limit they obtain non-zero values at precisely the same critical coupling $U_c$.
Hence in total we observe a semimetal-antiferromagnetic Mott insulator (SM-AFMI) transition which falls into the Gross-Neveu-Heisenberg universality class. Up to date summaries of the critical parameters can be found in~\cite{lattice_pos-21,Janssen_2022}. 

\section{Fermionic Projected Entangled Pair States}\label{sec:peps}

Away from half filling, i.e.\ at non-zero chemical potential $\mu$, the so-called fermion sign problem emerges. It manifests itself in a `probability density'
\begin{align}
	p[\phi] &\propto \det\left(M[\phi,\mu]M[\phi,-\mu]^\dagger\right)
\end{align}
that is no longer positive semi-definite. Thus, Monte-Carlo simulations cannot be performed without additional considerations.

The most straight-forward approach to restore stochastic tractability is the reweighting technique where the complex phase $\eto{\im\theta}$ of the weight $p[\phi]$ is treated as part of the observable. That is, the HMC simulation proceeds as usual, but with the probability density $|p[\phi]|$, and the expectation value of an observable $O$ is obtained via
\begin{align}
	\erwartung{O}_p &= \frac{\erwartung{O\eto{\im\theta}}_{|p|}}{\erwartung{\eto{\im\theta}}_{|p|}}\,.
\end{align}
This method, however, quickly becomes very unstable when the statistical power $\erwartung{\eto{\im\theta}}_{|p|}$ is small.

There is a large variety of alternative algorithms avoiding or alleviating the sign problem. They include, but are by no means restricted to, simulations close to the Lefschetz thimbles using holomorphic flow or machine learning~\cite{leveragingML,complexNN,Ulybyshev:2019fte} and density of states methods~\cite{PhysRevD.102.054502,real_time}.

In the rest of this section we will focus on Tensor Network (TN) simulations that do not have a sign problem at all because they do not rely on probability sampling. More precisely, we use fermionic Projected Entangled Pair States (PEPS)~\cite{PEPS_original_bMPS,tensor_fermions} closely following Ref.~\cite{my_tensor_networks}.

\subsection{Formalism}
In order to get an intuition for the TN approach, it is most instructive to start in $d=1$ dimension with so-called matrix product states (MPS). They can be derived using successive singular value decompositions (SVD) on a mixed quantum state
\begin{align}
	\Ket{\psi} &= \sum_{s_1}\sum_{s_2}\cdots\sum_{s_N}A_{s_1,s_2,\dots,s_N}\Ket{s_1}\otimes\Ket{s_2}\otimes\cdots \otimes \Ket{s_N}\\
	&= \sum_{s_1}\sum_{s_2}\cdots\sum_{s_N}A_{s_1;\alpha_1}^1A_{s_2;\alpha_1,\alpha_2}^2\cdots A_{s_N;\alpha_{N-1}}^N\Ket{s_1}\otimes\Ket{s_2}\otimes\cdots \otimes \Ket{s_N}\label{eq:svd_mps}
\end{align}
composed from local finite-dimensional degrees of freedom $\Ket{s_i}$. Now the number of parameters in the rank-$N$ tensor $A_{s_1,s_2,\dots,s_N}$ grows exponential in $N$. The constituent tensors $A_{s_i;\alpha_{i-1},\alpha_i}$ on the other hand can be truncated to some bond dimension $D$ so that relation~\eqref{eq:svd_mps} is not exact any more, but the number of parameters grows merely linearly in $N$. In practice moderate bond dimensions (i.e.\ those tractable on a computer) often lead to good approximations.

\begin{figure}[th]
	\centering
	\hfill
	\raisebox{2\baselineskip}{
\begin{tikzpicture}
	\def\Lx{3}
	\def\Ly{4}
	\def\xStep{\xDist}
	\def\yStep{-\yDist}
	\def\physLength{-\yExt}
	\def\xShift{\xStep / \Ly} %
	\node (origin) at (- \xShift*3, \yStep) {};
	\foreach \x in {1,...,\Lx}
		\foreach \y in {1,...,\Ly}
			{
				\node[ket] (ten_\x_\y) at (\xStep*\x - \xShift * \y, \yStep*\y) {};
				\node (physEnd_\x_\y) at (\xStep*\x - \xShift * \y, {\yStep*\Ly + \physLength}) {};
			}
		\draw[parity] (ten_1_\Ly) -- (-\xShift,{\yStep*\Ly + \physLength});
		\foreach \x in {1,...,\Lx}
			\foreach \y in {1,...,\Ly}
			{
				\draw[physical,name path={phys_\x_\y}] (ten_\x_\y) -- (physEnd_\x_\y);
			}
		\foreach \x in {2,...,\Lx}
			\foreach \y in {1,...,\Ly}
			{
				\ifthenelse{\intcalcMod{\x+\y}{2}=0}{
				}{
					\pgfmathtruncatemacro\lastx{\x-1}
					\pgfmathtruncatemacro\lasty{\y-1}
					\draw[name path={xlink_\lastx_\y}] (ten_\lastx_\y) -- (ten_\x_\y);
					\ifthenelse{\y>1}{ %
						\foreach \physy in {1,...,\lasty}
						{
							\path [name intersections={of={xlink_\lastx_\y} and {phys_\lastx_\physy},by={swap_\lastx_\y_\physy}}];
							\node[swap]  at ({swap_\lastx_\y_\physy}) {};
						}
					}
				}
			}
		\foreach \x in {1,...,\Lx}
			\foreach \y in {2,...,\Ly}
			{
				\pgfmathtruncatemacro\lasty{\y-1}
				\draw (ten_\x_\lasty) -- (ten_\x_\y);
			}
\end{tikzpicture} %
}
	\hfill
\begingroup%
  \makeatletter%
  \providecommand\color[2][]{%
    \errmessage{(Inkscape) Color is used for the text in Inkscape, but the package 'color.sty' is not loaded}%
    \renewcommand\color[2][]{}%
  }%
  \providecommand\transparent[1]{%
    \errmessage{(Inkscape) Transparency is used (non-zero) for the text in Inkscape, but the package 'transparent.sty' is not loaded}%
    \renewcommand\transparent[1]{}%
  }%
  \providecommand\rotatebox[2]{#2}%
  \newcommand*\fsize{\dimexpr\f@size pt\relax}%
  \newcommand*\lineheight[1]{\fontsize{\fsize}{#1\fsize}\selectfont}%
  \ifx\svgwidth\undefined%
    \setlength{\unitlength}{126.45096606bp}%
    \ifx\svgscale\undefined%
      \relax%
    \else%
      \setlength{\unitlength}{\unitlength * \real{\svgscale}}%
    \fi%
  \else%
    \setlength{\unitlength}{\svgwidth}%
  \fi%
  \global\let\svgwidth\undefined%
  \global\let\svgscale\undefined%
  \makeatother%
  \begin{picture}(1,0.62481354)%
    \lineheight{1}%
    \setlength\tabcolsep{0pt}%
    \put(0,0){\includegraphics[width=\unitlength,page=1]{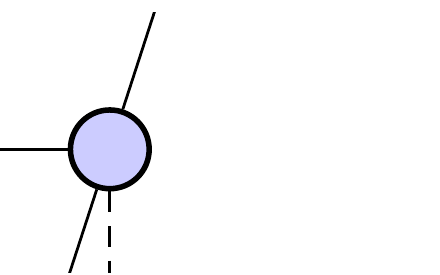}}%
    \put(5.7584102,-0.66934156){\color[rgb]{0,0,0}\makebox(0,0)[lt]{\lineheight{1.25}\smash{\begin{tabular}[t]{l}4/4\end{tabular}}}}%
    \put(0.19720186,0.23816808){\color[rgb]{0,0,0}\makebox(0,0)[lt]{\lineheight{1.25}\smash{\begin{tabular}[t]{l}$A$\end{tabular}}}}%
    \put(0.26983199,0.01859663){\color[rgb]{0,0,0}\makebox(0,0)[lt]{\lineheight{1.25}\smash{\begin{tabular}[t]{l}$s$\end{tabular}}}}%
    \put(0.35739668,0.52867944){\color[rgb]{0,0,0}\makebox(0,0)[lt]{\lineheight{1.25}\smash{\begin{tabular}[t]{l}$\alpha_i$\end{tabular}}}}%
    \put(0.00408744,0.33411406){\color[rgb]{0,0,0}\makebox(0,0)[lt]{\lineheight{1.25}\smash{\begin{tabular}[t]{l}$\alpha_j$\end{tabular}}}}%
    \put(0.02075149,0.03184634){\color[rgb]{0,0,0}\makebox(0,0)[lt]{\lineheight{1.25}\smash{\begin{tabular}[t]{l}$\alpha_k$\end{tabular}}}}%
  \end{picture}%
\endgroup%
 	\hfill
	\caption{State of a PEPS for a 3x4 fermionic honeycomb lattice (left) and single tensor representation (right). Description of symbology (see text for more details) -- circles: PEPS tensors; dashed lines: physical indices; solid lines: internal indices; dotted line: parity index; diamonds: swap gates.}\label{fig:peps_structure}
\end{figure}
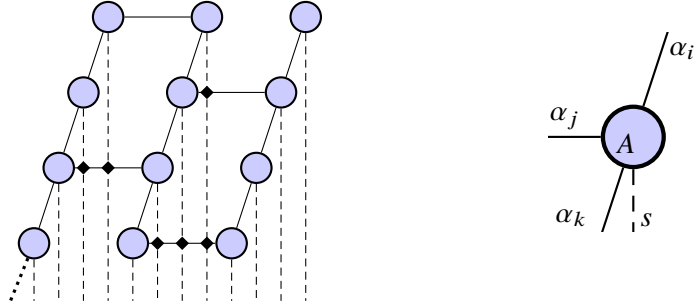

\begin{figure}[th]
	\centering
\begingroup%
  \makeatletter%
  \providecommand\color[2][]{%
    \errmessage{(Inkscape) Color is used for the text in Inkscape, but the package 'color.sty' is not loaded}%
    \renewcommand\color[2][]{}%
  }%
  \providecommand\transparent[1]{%
    \errmessage{(Inkscape) Transparency is used (non-zero) for the text in Inkscape, but the package 'transparent.sty' is not loaded}%
    \renewcommand\transparent[1]{}%
  }%
  \providecommand\rotatebox[2]{#2}%
  \newcommand*\fsize{\dimexpr\f@size pt\relax}%
  \newcommand*\lineheight[1]{\fontsize{\fsize}{#1\fsize}\selectfont}%
  \ifx\svgwidth\undefined%
    \setlength{\unitlength}{329.55295907bp}%
    \ifx\svgscale\undefined%
      \relax%
    \else%
      \setlength{\unitlength}{\unitlength * \real{\svgscale}}%
    \fi%
  \else%
    \setlength{\unitlength}{\svgwidth}%
  \fi%
  \global\let\svgwidth\undefined%
  \global\let\svgscale\undefined%
  \makeatother%
  \begin{picture}(1,0.3113469)%
    \lineheight{1}%
    \setlength\tabcolsep{0pt}%
    \put(0,0){\includegraphics[width=\unitlength,page=1]{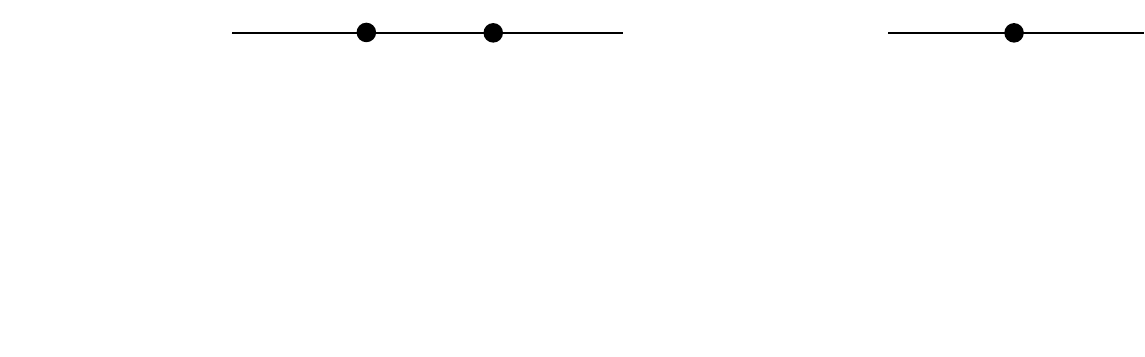}}%
    \put(0.63319458,0.27445981){\color[rgb]{0,0,0}\makebox(0,0)[lt]{\lineheight{1.25}\smash{\begin{tabular}[t]{l}$=$\end{tabular}}}}%
    \put(0,0){\includegraphics[width=\unitlength,page=2]{MPS-vs-TN_contraction.pdf}}%
    \put(0.63319458,0.08784324){\color[rgb]{0,0,0}\makebox(0,0)[lt]{\lineheight{1.25}\smash{\begin{tabular}[t]{l}$=$\end{tabular}}}}%
    \put(0,0){\includegraphics[width=\unitlength,page=3]{MPS-vs-TN_contraction.pdf}}%
    \put(-0.00403008,0.27445981){\color[rgb]{0,0,0}\makebox(0,0)[lt]{\lineheight{1.25}\smash{\begin{tabular}[t]{l}$d=1$\end{tabular}}}}%
    \put(-0.00403008,0.08784324){\color[rgb]{0,0,0}\makebox(0,0)[lt]{\lineheight{1.25}\smash{\begin{tabular}[t]{l}$d>1$\end{tabular}}}}%
  \end{picture}%
\endgroup%
 	\caption{Visualisation of two tensor contractions in one and two spatial dimensions respectively. In 1D the contraction leads to a self-similar object (matrix-matrix multiplication yielding a matrix) whereas dimensions larger than one the contraction results in a tensor of higher rank than each of its constituents (contracting two rank-3 tensors yielding a rank-4 tensor).}\label{fig:more_is_different}
\end{figure}

The generalisation to more than one spatial dimension is straight forward in this formalism. In $d=2$ dimensions the object thus obtained is a PEPS and it can be visualised as in figure~\ref{fig:peps_structure}. We remark at this point that even though this is not challenging mathematically, higher dimensions fundamentally increase computational complexity. The crucial difference is that in $d>1$ some tensors have 3 or more internal links and the contractions of two such objects results in a tensor of even higher rank (see fig.~\ref{fig:more_is_different}). Therefore additional truncations are required when contracting a PEPS. Here we use the boundary MPS approach where the PEPS is contracted line by line and the links between the tensors on the boundary line are truncated to the dimension $\chi$. In literature usually $\chi\simeq D^2$ is chosen, however we find that $\chi\le 3D$ is enough in most cases leading to a significant speed up.

Another challenge is presented by the fermionic anti-commutation relations. They translate to non-trivial behaviour whenever lines (or links) of the PEPS cross. We incorporate the property that any even number of fermions commutes with any number while two odd numbers anti-commute by introducing even- and odd- parity sectors and the swap gate
\begin{align*}
	&\qquad\;\;\overbrace{\hphantom{1\;\; \dots\ 1}}^\text{even}\;\;\;\overbrace{\hphantom{-1\; \dots\ -1}}^\text{odd}\\
	S &= \matr{rcrrcr}{1&\dots&1&1&\dots&1\\\vdots&\ddots&\vdots&\vdots&\ddots&\vdots\\1&\dots&1&1&\dots&1\\1&\dots&1&-1&\dots&-1\\\vdots&\ddots&\vdots&\vdots&\ddots&\vdots\\1&\dots&1&-1&\dots&-1}
	\!\!\!\!\!\!\!\!\!
	\begin{array}{l}
		\left.
		\begin{array}{l}
			\vphantom{1}\\\vphantom{\vdots}\\\vphantom{1}
		\end{array}
		\right\}\text{even}\\						
		\left.
		\begin{array}{l}
			\vphantom{1}\\\vphantom{\vdots}\\\vphantom{1}
		\end{array}
		\right\}\text{odd}
	\end{array}
\end{align*}
that has to be inserted at every crossing of two links (diamonds in fig.~\ref{fig:peps_structure}). The overall parity of the system is a conserved quantity and can be fixed by an external parity link as shown in the bottom left corner of figure~\ref{fig:peps_structure}.

\subsection{Imaginary time evolution}
In contrast to the canonical approach chosen for the HMC algorithm in \Cref{sec:hmc}, we do not simulate PEPS at finite temperature. Instead we perform a ground state search for which the TN ansatz is much better suited. To this end we evolve a random initial state in a given parity sector in imaginary time until convergence is reached. The time steps are decreased simultaneously, so that time discretisation artifacts can be eliminated completely.

Details on the time step reduction scheme are provided in Section~III.B of Ref.~\cite{my_tensor_networks}. Crucially, we can monitor the rate of convergence by means of the cheap energy estimator
\begin{align}
	E &\approx - \frac{1}{\delta t} \ln \sqrt{\frac{\Braket{\Psi'|\Psi'}}{\Braket{\Psi|\Psi}}}\,,\label{eq:direct_estim}
\end{align}
where $\Ket{\Psi}$ is the state before and $\Ket{\Psi'}$ the state after a single imaginary time evolution step of length~$\delta t$. The change of the norm here can simply be calculated as the product of norm changes of all the individual local tensors, thus no full contraction of the network is needed.

\begin{figure}[ht]
	\centering
	\raisebox{2.6\height}{
		\resizebox{.45\textwidth}{!}{{\normalsize\begin{tikzpicture}
	\node[ket] (triadp) at (0, 0) {};
	\node[ket] (triadm) at \rightOf{triadp} {};
	\draw (triadp) -- (triadm) node[nodeInvisible,midway] (S) {};
	\draw[wiggly] \lineL{triadp};
	\draw[wiggly] \lineR{triadm};
	\node[gate] (expGate) at \belowOf{S} {$\exp(-\tau H_i)$};
	\draw[physical] \connectD{triadp}{expGate};
	\draw[physical] \connectD{triadm}{expGate};
	\draw[physical] \gateLD{expGate};
	\draw[physical] \gateRD{expGate};
\end{tikzpicture}
\begin{tikzpicture}
	\draw[-latex] (0,0) -- (1.3cm,0) node[midway,above] {\scriptsize truncate};
\end{tikzpicture}
\begin{tikzpicture}
	\node[ket] (triadp) at (0, 0) {};
	\node[ket] (triadm) at \rightOf{triadp} {};
	\draw (triadp) -- (triadm) {};
	\draw[wiggly] \lineL{triadp};
	\draw[wiggly] \lineR{triadm};
	\draw[physical] \lineD{triadp};
	\draw[physical] \lineD{triadm};
\end{tikzpicture}}}}
	\hfill
	\resizebox{.45\textwidth}{!}{{\normalsize\input{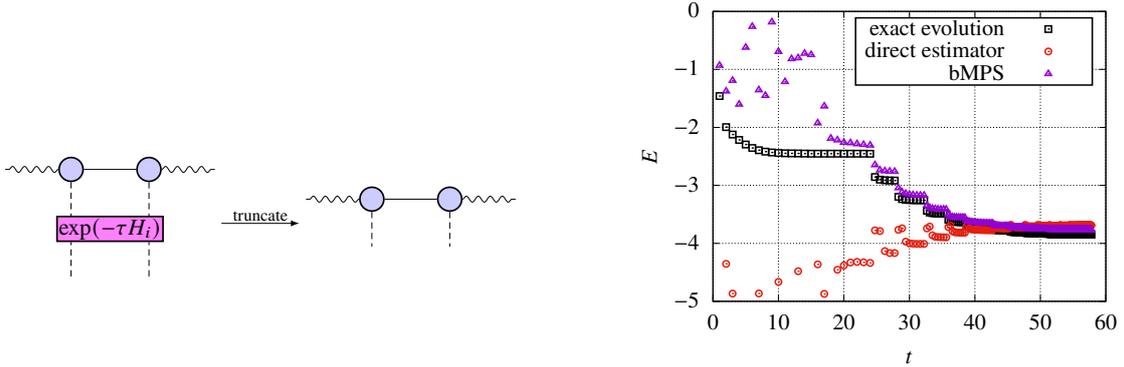}}}
	\caption{Left: Singe step of the Simple Update algorithm, a gate is applied locally and the result is truncated.
		Right: Imaginary time evolution with $\kappa=1$, $U=4$, $\mu=B=\num{0.1}$ on the $2\times 4$-lattice with $D=8$ and odd parity. Energies calculated using boundary MPS (bMPS) and the direct estimator from eq.~\eqref{eq:direct_estim}. As a reference we provide the exact imaginary time evolution of the state vector obtained via full contraction of the initial PEPS.}\label{fig:convergence}
\end{figure}

We use the Simple Update (SU) scheme for the imaginary time evolution that has the same advantage of not requiring a full TN contraction~\cite{Schneider_2022}. It consists of local updates applying so-called gates, i.e.\ Trotter-decomposed components of the time evolution operator, to each pair of sites successively. Such a SU step is visualised in the left panel of figure~\ref{fig:convergence}. Only a single contraction of the complete TN with boundary MPS is required at the end of the time evolution to obtain expectation values for observables.
Since full contractions scale at least as $\ordnung{D^7}$ with the bond dimension while SU only scales as $\ordnung{D^4}$, the runtime is mostly governed by the single contraction at the end of the simulation and the algorithm is by several orders of magnitude faster than the alternative Full Update.

A typical convergence plot with appropriately tuned parameters is shown on the right of figure~\ref{fig:convergence}. For this small system size a comparison with the exact evolution of the full state is possible and we find good agreement between the exact results and the true boundary MPS estimator. Both converge to the correct ground state energy simultaneously approaching the infinite and continuous time limits. The direct estimator from equation~\eqref{eq:direct_estim} is inaccurate, but it captures the convergence behaviour qualitatively.

\subsection{Results}
\begin{figure}[ht]
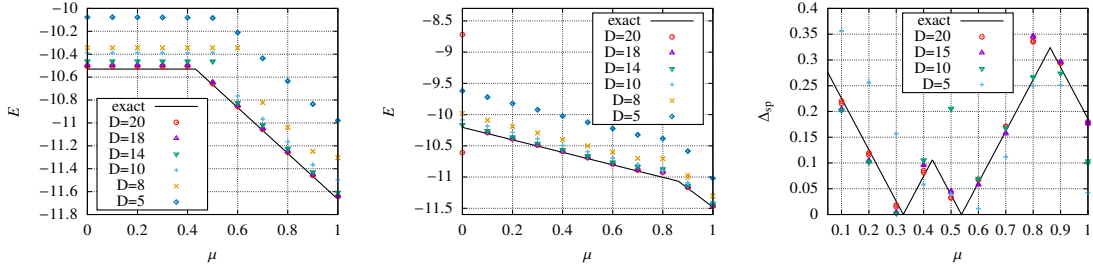

	\centering
	\resizebox{.32\textwidth}{!}{{\normalsize\input{plots/even_energy_of_mu}}}
	\resizebox{.32\textwidth}{!}{{\normalsize\input{plots/odd_energy_of_mu}}}
	\resizebox{.32\textwidth}{!}{{\normalsize\input{plots/gap_of_mu}}}
	\caption{Energies of the $3\times 4$ hexagonal lattice with $\kappa=1$, $U=2$ and $B=0$ at different values of $\mu$. Duplicate points correspond to $\chi=2D$ and $\chi=3D$. Left: Even parity; center: Odd parity; right: Energy gap between even and odd parity sectors.}\label{fig:small_results}
\end{figure}

Having explained the TN formalism, we have to test its usefulness in a case where the HMC algorithm fails. We therefore simulated the $3\times4$ honeycomb lattice (the largest lattice we could solve with exact diagonalisation for benchmarking) at finite chemical potential where the HMC suffers from a very severe sign problem. The results can be found in figure~\ref{fig:small_results} for the ground states of both parity sectors and the modulus of their difference, i.e.\ the single particle gap. They do not only converge well with the bond dimension $D$, the simulations also proved significantly less compute intensive than the exact diagonalisation. The only regions of bad convergence are close to the cross overs from one ground state to another (kinks in the solid lines) because the ground states are ambiguous in these regions.

\begin{figure}[ht]
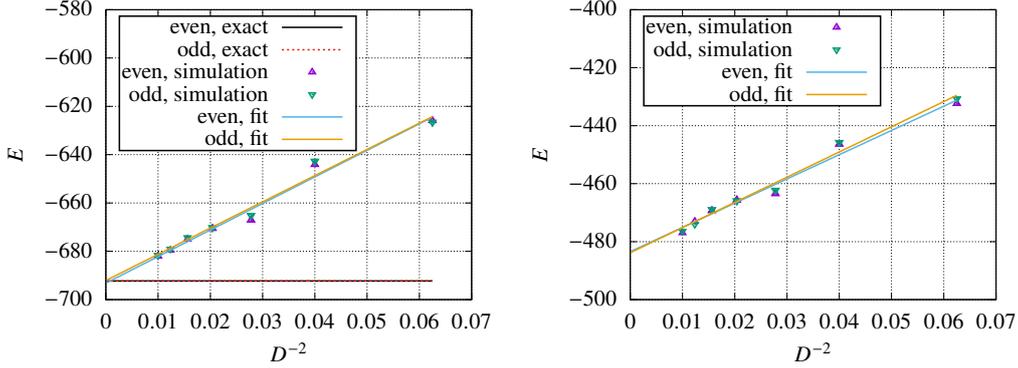

	\centering
	\resizebox{.45\textwidth}{!}{{\normalsize\input{plots/energy_of_inv_D_sq_mu05}}}
	\resizebox{.45\textwidth}{!}{{\normalsize\input{plots/energy_of_inv_D_sq_mu05_U2}}}
	\caption{Energies with finite chemical potential ($\kappa=1$, $\mu=0.5$, $B=0$) for the $30\times 15$-lattice against the inverse squared bond dimension. Duplicate points correspond to $\chi=2D$ and $\chi=3D$. Left: non-interacting, i.e.\ $U=0$; right: $U=2$.}\label{fig:large_results}
\end{figure}

\Cref{fig:large_results} demonstrates that the TN method scales to lattice sizes far beyond exact diagonalisability. On the left we show the non-interacting case of the $30\times15$ honeycomb lattice away from half filling. The results obtained from the PEPS ground state search converge with $\sim D^{-2}$ towards the correct value. The right hand side plot of figure~\ref{fig:large_results} shows the first prediction for an interacting lattice of this size away from half filling. We chose $U/\kappa=2$, $\mu/\kappa=\num{0.5}$ and obtained the even and odd ground state energies $E_\text{even} = \num{-483.5(14)}$ and $E_\text{odd} = \num{-483.8(12)}$ respectively.

Let us finally remark that our TN simulations did not violate the no-free-lunch theorem yet. The first reason is that the ground state energy is an extensive quantity. This means that even though we can reliably estimate it from PEPS ground state search with affordable computational effort and acceptable relative precision, it is unfeasible to keep the absolute error at a constant level with growing system size. Therefore intensive quantities like the single particle energy gap cannot be resolved for large systems. Usually intensive observables carry most interesting physical information and it is a challenge to extract as much physical insight as possible from the available extensive observables.

\begin{figure}[ht]
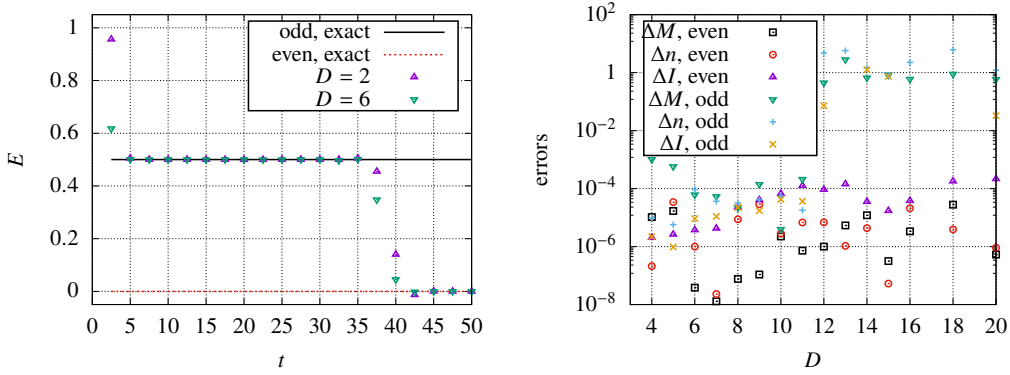

	\centering
	\resizebox{.45\textwidth}{!}{{\normalsize\input{plots/Hubb_K0_U1_Mu0_B0_L3x4_odd_simple}}}
	\resizebox{.45\textwidth}{!}{{\normalsize\input{plots/deviations_of_D_U3}}}
	\caption{Left: Energy during the imaginary time evolution with $U=1$ and $\kappa=\mu=B=0$ on the $3\times 4$-lattice with odd parity. The energies have been calculated using boundary MPS (bMPS) for Simple Update.
	Right: Standard deviation of the norm $\Delta I$ and deviations of the magnetization $M$ and the particle number $n$ from the exact value (see Ref.~\cite{my_tensor_networks} for more details).  $3\times 4$ hexagonal lattice with $\kappa=1$, $U=3$ and $\mu=B=0$ for different bond dimensions $D$ using $\chi=3D$.}\label{fig:jump_down}
\end{figure}

Moreover so far we only showed parameter regions with well behaved convergence. We find, however, that stable convergence is not guaranteed in the case of a large gap between the ground state of the excited parity sector (usually odd parity) and the true ground state. An extreme case is presented on the left hand side of figure~\ref{fig:jump_down} where the correct ground state of the odd parity sector is approached at first, but then numerical instabilities enforce a jump into the forbidden even parity sector with its lower lying ground state. This implies that only the results of a global ground state search can be fully trusted while results in a particular parity sector might be deceptive.

It is important to note that these simulations fail `gracefully' in the sense that a failure can be clearly identified even if the correct result is unknown. For instance the norm of a state can be calculated in our framework, but of course the state should be normalised to start with. Large standard deviations of the norm $\Delta I$ therefore indicate numerical errors. We plot several observables of this type in the right panel of figure~\ref{fig:jump_down} and the region where the simulations fail (odd parity, $D\ge12$) is clearly visible. 

\section{Conclusion}\label{sec:conclusion}

In this proceeding we have presented two fundamentally different algorithms for the simulation of quantum mechanical systems applied to the Hubbard model on the honeycomb lattice. On the one hand the Hybrid Monte Carlo (HMC) algorithm has been explained in \Cref{sec:hmc} together with a summary of the quantum phase transition we could extract relying on data from HMC simulations. The HMC algorithm has been well established in the lattice field theory community for more than three decades by now and can be considered the default approach, the `work horse'. Fermionic Projected Entangled Pair States (PEPS) on the other hand are a rather young variety of Tensor Networks (TN) and not even in their teens yet. In \Cref{sec:peps} we recalled the current state of fermionic PEPS ground state search simulations.
Let us now provide a more detailed analysis of the respective advantages and disadvantages of the two computational methods.

As to date the HMC algorithm is applicable to very large systems with $\ordnung{10^4}$ spatial lattice sites while fermionic PEPS do not exceed $\ordnung{10^3}$ sites. Moreover fermionic PEPS have to be formulated with open boundary conditions in order to apply the boundary Matrix Product States (MPS) contraction method. For the HMC arbitrary boundary conditions can be chosen, in particular periodic boundaries guarantee faster convergence towards the thermodynamic limit.

In all these considerations the HMC algorithms is compute and band width bound, whereas TN are almost entirely limited by sheer memory requirement.

It is also noteworthy that the HMC simulations on a $d+1$ dimensional space-time lattice have to be extrapolated to the continuous time limit and are restricted to finite temperature calculations. Fermionic PEPS on the other hand easily approach the continuum limit by successive step size reduction and their ground state search produces only zero temperature results.

A very serious disadvantage of the TN method lies in the poorly controlled convergence in the bond dimension $D$ and the lack of a reliable means to estimate the error of the final results.

Excited state calculations are challenging with both algorithms. While the HMC results require high precision data and complicated generalised eigenvalue type analyses~\cite{prony_gevm}, excited TN states can be obtained by first finding the ground state and then projecting it out in a next iteration of the imaginary time evolution. This projection quickly becomes numerically unstable. Fermionic PEPS allow for one exception since they give access to the even and odd parity ground states independently. Some care, however, is called for in this case as well because stable convergence is not guaranteed in systems with a large gap.

Of course, the crucial advantage of the TN approach over stochastic methods like the HMC lies in the total absence of the fermionic sign problem allowing for simulations of otherwise totally unreachable regions of the phase space.

All in all the two algorithms are not truly competing. Rather they complement each other allowing for different types of simulations, the HMC being ideal for large scale computations near half filling and fermionic PEPS well suited away from half filling.

\section*{Acknowledgements}

We thank Evan Berkowitz, Stefan Krieg, Timo Lähde, Tom Luu, Marcel Rodekamp and Carsten Urbach for helpful discussions on the Hubbard model and implementation details of the HMC algorithm. We also thank Manuel Schneider and Karl Jansen for the fruitful collaboration on tensor networks.
This work was funded, in part, through financial support from the Deutsche Forschungsgemeinschaft (Sino-German CRC 110 and SFB TRR-55)
as well as the STFC Consolidated Grant ST/T000988/1.
We gratefully acknowledge the Computer Center at DESY Zeuthen for the compute time,
the computing time granted through JARA-HPC on the supercomputer JURECA~\cite{jureca} at Forschungszentrum J\"ulich,
and the time on DEEP~\cite{DEEP}, an experimental modular supercomputer at the J\"ulich Supercomputing Centre.

\FloatBarrier
\bibliographystyle{JHEP}
\bibliography{cns}

\providecommand{\href}[2]{#2}\begingroup\raggedright\begin{thebibliography}{10}

\bibitem{GeimNovoselovReview}
A.K.~Geim and K.S.~Novoselov, \emph{The rise of graphene}, {\emph{Nat Mater}
  {\bfseries 6} (2007) 183}.

\bibitem{CastroNeto2009}
A.H.~Castro~Neto, F.~Guinea, N.M.R.~Peres, K.S.~Novoselov and A.K.~Geim,
  \emph{The electronic properties of graphene},
  \href{https://doi.org/10.1103/RevModPhys.81.109}{\emph{Rev. Mod. Phys.}
  {\bfseries 81} (2009) 109}.

\bibitem{Luu:2015gpl}
T.~Luu and T.A.~L{\"a}hde, \emph{{Quantum Monte Carlo Calculations for Carbon
  Nanotubes}}, \href{https://doi.org/10.1103/PhysRevB.93.155106}{\emph{Phys.
  Rev.} {\bfseries B93} (2016) 155106}.

\bibitem{Assaad:2013xua}
F.F.~Assaad and I.F.~Herbut, \emph{{Pinning the order: the nature of quantum
  criticality in the Hubbard model on honeycomb lattice}},
  \href{https://doi.org/10.1103/PhysRevX.3.031010}{\emph{Phys. Rev.} {\bfseries
  X3} (2013) 031010}.

\bibitem{Otsuka:2015iba}
Y.~Otsuka, S.~Yunoki and S.~Sorella, \emph{{Universal Quantum Criticality in
  the Metal-Insulator Transition of Two-Dimensional Interacting Dirac
  Electrons}}, \href{https://doi.org/10.1103/PhysRevX.6.011029}{\emph{Phys.
  Rev.} {\bfseries X6} (2016) 011029}.

\bibitem{Kotov2012}
V.N.~Kotov, B.~Uchoa, V.M.~Pereira, F.~Guinea and A.H.~Castro~Neto,
  \emph{{Electron-Electron Interactions in Graphene: Current Status and
  Perspectives}}, \href{https://doi.org/10.1103/RevModPhys.84.1067}{\emph{Rev.
  Mod. Phys.} {\bfseries 84} (2012) 1067}.

\bibitem{more_observables}
J.~Ostmeyer, E.~Berkowitz, S.~Krieg, T.A.~Lähde, T.~Luu and C.~Urbach,
  \emph{{The Antiferromagnetic Character of the Quantum Phase Transition in the
  Hubbard Model on the Honeycomb Lattice}},
  \href{https://doi.org/10.1103/PhysRevB.104.155142}{\emph{Phys. Rev. B}
  {\bfseries 104} (2021) 155142}.

\bibitem{my_ising}
J.~Ostmeyer, E.~Berkowitz, T.~Luu, M.~Petschlies and F.~Pittler, \emph{{The
  Ising model with Hybrid Monte Carlo}},
  \href{https://doi.org/10.1016/j.cpc.2021.107978}{\emph{Computer Physics
  Communications} {\bfseries 265} (2021) 107978}.

\bibitem{Brower:2012zd}
R.~Brower, C.~Rebbi and D.~Schaich, \emph{{Hybrid Monte Carlo simulation on the
  graphene hexagonal lattice}},
  \href{https://doi.org/10.22323/1.139.0056}{\emph{PoS} {\bfseries LATTICE2011}
  (2011) 056} [\href{https://arxiv.org/abs/1204.5424}{{\ttfamily 1204.5424}}].

\bibitem{Blankenbecler:1981jt}
R.~Blankenbecler, D.J.~Scalapino and R.L.~Sugar, \emph{{Monte Carlo
  Calculations of Coupled Boson - Fermion Systems. 1.}},
  \href{https://doi.org/10.1103/PhysRevD.24.2278}{\emph{Phys. Rev.} {\bfseries
  D24} (1981) 2278}.

\bibitem{acceleratingHMC}
S.~{Krieg}, T.~{Luu}, J.~{Ostmeyer}, P.~{Papaphilippou} and C.~{Urbach},
  \emph{{Accelerating Hybrid Monte Carlo simulations of the Hubbard model on
  the hexagonal lattice}},
  \href{https://doi.org/10.1016/j.cpc.2018.10.008}{\emph{Computer Physics
  Communications} (2018) }.

\bibitem{Smith:2014tha}
D.~Smith and L.~von Smekal, \emph{{Monte-Carlo simulation of the tight-binding
  model of graphene with partially screened Coulomb interactions}},
  \href{https://doi.org/10.1103/PhysRevB.89.195429}{\emph{Phys. Rev.}
  {\bfseries B89} (2014) 195429}.

\bibitem{Wynen:2018ryx}
J.-L.~Wynen, E.~Berkowitz, C.~K{\"o}rber, T.A.~L{\"a}hde and T.~Luu,
  \emph{{Avoiding Ergodicity Problems in Lattice Discretizations of the Hubbard
  Model}}, \href{https://doi.org/10.1103/PhysRevB.100.075141}{\emph{Phys. Rev.}
  {\bfseries B100} (2019) 075141}.

\bibitem{Buividovich:2018yar}
P.~Buividovich, D.~Smith, M.~Ulybyshev and L.~von Smekal, \emph{{Hybrid Monte
  Carlo study of competing order in the extended fermionic Hubbard model on the
  hexagonal lattice}},
  \href{https://doi.org/10.1103/PhysRevB.98.235129}{\emph{Phys. Rev. B}
  {\bfseries 98} (2018) 235129}.

\bibitem{prony_gevm}
M.~Fischer, B.~Kostrzewa, J.~Ostmeyer, K.~Ottnad, M.~Ueding and C.~Urbach,
  \emph{{On the generalised eigenvalue method and its relation to Prony and
  generalised pencil of function methods}},
  \href{https://doi.org/10.1140/epja/s10050-020-00205-w}{\emph{The European
  Physical Journal A} {\bfseries 56} (2020) }.

\bibitem{hadron}
B.~Kostrzewa, J.~Ostmeyer, M.~Ueding and C.~Urbach, \emph{{hadron: Analysis
  Framework for Monte Carlo Simulation Data in Physics}}, 2020.

\bibitem{R_language}
{R Core Team}, \emph{{R: A Language and Environment for Statistical
  Computing}}.
\newblock R Foundation for Statistical Computing, Vienna, Austria, 2018.

\bibitem{semimetalmott}
J.~Ostmeyer, E.~Berkowitz, S.~Krieg, T.A.~L\"ahde, T.~Luu and C.~Urbach,
  \emph{{Semimetal--Mott insulator quantum phase transition of the Hubbard
  model on the honeycomb lattice}},
  \href{https://doi.org/10.1103/PhysRevB.102.245105}{\emph{Phys. Rev. B}
  {\bfseries 102} (2020) 245105}.

\bibitem{lattice_pos-21}
J.~Ostmeyer, E.~Berkowitz, S.~Krieg, T.~Lahde, T.~Luu and C.~Urbach, \emph{{The
  Semimetal-Antiferromagnetic Mott Insulator Quantum Phase Transition of the
  Hubbard Model on the Honeycomb Lattice}},
  \href{https://doi.org/10.22323/1.396.0303}{\emph{PoS} {\bfseries LATTICE2021}
  (2022) 303}.

\bibitem{Janssen_2022}
K.~Ladovrechis, S.~Ray, T.~Meng and L.~Janssen, \emph{{Gross-Neveu-Heisenberg
  criticality from $2+\varepsilon$ expansion}}, {\emph{arXiv e-prints} (2022) }
  [\href{https://arxiv.org/abs/2209.02734}{{\ttfamily 2209.02734}}].

\bibitem{leveragingML}
J.-L.~Wynen, E.~Berkowitz, S.~Krieg, T.~Luu and J.~Ostmeyer, \emph{{Machine
  learning to alleviate Hubbard-model sign problems}},
  \href{https://doi.org/10.1103/PhysRevB.103.125153}{\emph{Phys. Rev. B}
  {\bfseries 103} (2021) 125153}.

\bibitem{complexNN}
M.~Rodekamp, E.~Berkowitz, C.~Gäntgen, S.~Krieg, T.~Luu and J.~Ostmeyer,
  \emph{{Mitigating the Hubbard Sign Problem with Complex-Valued Neural
  Networks}}, \href{https://doi.org/10.1103/PhysRevB.106.125139}{\emph{Phys.
  Rev. B} {\bfseries 106} (2022) 125139}
  [\href{https://arxiv.org/abs/2203.00390}{{\ttfamily 2203.00390}}].

\bibitem{Ulybyshev:2019fte}
M.~Ulybyshev, C.~Winterowd and S.~Zafeiropoulos, \emph{{Lefschetz thimbles
  decomposition for the Hubbard model on the hexagonal lattice}},
  \href{https://doi.org/10.1103/PhysRevD.101.014508}{\emph{Phys. Rev. D}
  {\bfseries 101} (2020) 014508}
  [\href{https://arxiv.org/abs/1906.07678}{{\ttfamily 1906.07678}}].

\bibitem{PhysRevD.102.054502}
M.~K\"orner, K.~Langfeld, D.~Smith and L.~von Smekal, \emph{{Density of states
  approach to the hexagonal Hubbard model at finite density}},
  \href{https://doi.org/10.1103/PhysRevD.102.054502}{\emph{Phys.~Rev.~D}
  {\bfseries 102} (2020) 054502}
  [\href{https://arxiv.org/abs/2006.04607}{{\ttfamily 2006.04607}}].

\bibitem{real_time}
P.~Buividovich and J.~Ostmeyer, \emph{{Real Time Simulations of Quantum Spin
  Chains: Density-of-States and Reweighting approaches}}, {\emph{arXiv
  e-prints} (2022) } [\href{https://arxiv.org/abs/2209.13970}{{\ttfamily
  2209.13970}}].

\bibitem{PEPS_original_bMPS}
F.~{Verstraete} and J.I.~{Cirac}, \emph{{Renormalization algorithms for
  Quantum-Many Body Systems in two and higher dimensions}}, {\emph{arXiv
  e-prints} (2004) } [\href{https://arxiv.org/abs/cond-mat/0407066}{{\ttfamily
  cond-mat/0407066}}].

\bibitem{tensor_fermions}
P.~Corboz, R.~Or\'us, B.~Bauer and G.~Vidal, \emph{{Simulation of strongly
  correlated fermions in two spatial dimensions with fermionic projected
  entangled-pair states}},
  \href{https://doi.org/10.1103/PhysRevB.81.165104}{\emph{Phys. Rev. B}
  {\bfseries 81} (2010) 165104}.

\bibitem{my_tensor_networks}
M.~Schneider, J.~Ostmeyer, K.~Jansen, T.~Luu and C.~Urbach, \emph{{Simulating
  both parity sectors of the Hubbard Model with Tensor Networks}},
  \href{https://doi.org/10.1103/PhysRevB.104.155118}{\emph{Phys. Rev. B}
  {\bfseries 104} (2021) 155118}.

\bibitem{Schneider_2022}
M.~Schneider, \emph{{The Hubbard model on a honeycomb lattice with fermionic
  tensor networks}}, \href{https://doi.org/10.18452/25393}{\emph{(in press)}
  (2022) }.

\bibitem{jureca}
{J\"{u}lich Supercomputing Centre}, \emph{{JURECA: Modular supercomputer at
  J\"{u}lich Supercomputing Centre}},
  \href{https://doi.org/10.17815/jlsrf-4-121-1}{\emph{Journal of large-scale
  research facilities} {\bfseries 4} (2018) }.

\bibitem{DEEP}
N.~Eicker, T.~Lippert, T.~Moschny and E.~Suarez, \emph{{T}he {DEEP} {P}roject
  {A}n alternative approach to heterogeneous cluster-computing in the many-core
  era}, \href{https://doi.org/10.1002/cpe.3562}{\emph{Concurrency and
  computation} {\bfseries 28} (2016) 2394}.

\end{thebibliography}\endgroup

\end{document}